\documentclass[12pt]{article}

\usepackage{amssymb}
\usepackage{amsmath}
\usepackage{amsfonts}
\usepackage{graphicx}
\usepackage{subfigure}
\usepackage{color}
\usepackage{dcolumn}
\usepackage{hyperref}
\numberwithin{equation}{section}
\newcommand{\be}{\begin{equation}}
\newcommand{\ee}{\end{equation}}
\newcommand{\bea}{\begin{eqnarray}}
\newcommand{\eea}{\end{eqnarray}}
\newcommand{\nn}{\nonumber}

 \def\cB{{\cal B}} 
  \def\cF{{\cal F}}
 \def\cH{{\cal H}} 
  \def\cL{{\cal L}}
 \def\cN{{\cal N}} \def\cO{{\cal O}}

\def\a{\alpha}

  \def\D{\Delta}

\def\l{\lambda}

\def\s{\sigma}  
\def\t{\tau}
\def\th{\theta}

\newcommand{\prt}[1]{{\left( {#1} \right)}}
\newcommand{\eq}[1]{(\ref{#1})}
\newcommand{\la}[1]{\label{#1}}

\begin{document}

\begin{titlepage}
\thispagestyle{empty}

\hfill  WITS-CTP-113

\vspace{2cm}

\begin{center}
\font\titlerm=cmr10 scaled\magstep4 \font\titlei=cmmi10
scaled\magstep4 \font\titleis=cmmi7 scaled\magstep4 {
\Large{\textbf{The Imaginary Part of the Static Potential in Strongly Coupled Anisotropic Plasma }
\\}}
\vspace{1.5cm}
\noindent{{
Kazem Bitaghsir Fadafan$^{a}$\footnote{bitaghsir@shahroodut.ac.ir },
Dimitrios Giataganas$^{b,c~}$\footnote{dgiataganas@upatras.gr},
Hesam Soltanpanahi$^{c}$\footnote{hesam.soltanpanahisarabi@wits.ac.za }
}}\\
\vspace{0.8cm}

{\em
${^a}$Physics Department, Shahrood University of Technology, \\Shahrood, Iran\\}
\vspace*{.25cm}
{\em ${^b}$ Department of Engineering Sciences, University of Patras,\\
 26110 Patras, Greece\\}
\vspace*{.25cm}
{\em    ${^c}$
National Institute for Theoretical Physics,
School of Physics \\and Centre for Theoretical Physics,
University of the Witwatersrand,\\
Wits, 2050,
South Africa
}
\setcounter{footnote}{0}

\vspace*{.45cm}

\end{center}

\vskip 2em

\begin{abstract}
Using the gauge/gravity duality we study the imaginary part of the
static potential associated to the thermal width in finite temperature
strongly coupled anisotropic plasma. We firstly derive the
potential for a generic anisotropic background. Then we apply our formulas
to a theory where the anisotropy has been generated by a space dependent
axion term. We find that using our method there exist a peculiar
turning point in the imaginary part of the potential, similar to the
one appearing in the real part. The presence of anisotropy leads to decrease
of the imaginary potential, where larger decrease happens along the
anisotropic direction when the temperature is kept fixed. When the
entropy density is fixed, increase happens along the parallel direction
while along the transverse plane we observe a decrease. To estimate the thermal width
we use an approximate extrapolation beyond the turning point
and we find a decrease in presence of the anisotropy, independently of the comparison scheme used.
\end{abstract}

\end{titlepage}

\tableofcontents

\section{Introduction}
The experiments at Relativistic Heavy Ion Collisions (RHIC) and at
the Large Hadron Collider (LHC) have produced a strongly coupled
quark-gluon plasma (QGP) \cite{RHICa}. An interesting
property of the initial phase of the QGP are the anisotropies that
occur eg. \cite{anisofunction,Arnold:2005vb,Rebhan:2008uj}. One of these anisotropies caused by expansion of the plasma along the
collision axis, where unequal
pressures along the longitudinal and transverse directions occur.

Heavy quarkonium systems, like $J/\psi$ and excited states, have
turned out to provide extremely useful probes for QCD phenomenology
\cite{Brambilla:2004wf}. In general, the non-perturbative
determination of real and imaginary parts of heavy quarkonium  for
$1 < T/T_c < 2$ is particularly important in relation to the
fate of charmonium above the deconfined temperature $T_c$. It has been
found using a nonperturbative derivation of the Schr\"{o}dinger equation, that
close to $T_c$ the real part of the potential is not sensitive to the temperature
while the imaginary part is growing. As a result an increasing number of collisions
with the medium, may play more important role in  destabilization of heavy quarkonium than the screening effects \cite{Rothkopf:2011db,Gubler:2011ua}.

When the static potential
between heavy quarks is defined through the Schr\"{o}dinger equation at
high temperature in the weak coupling expansion, it has been shown
that the imaginary part appears naturally from the Landau damping of
approximately static fields
\cite{Laine:2006ns,Laine:2007gj,Beraudo:2007ky}.  In  \cite{Brambilla:2008cx} it was shown that apart from the Landau damping there is an additional  mechanism, the quark-antiquark color singlet to color octet thermal break up, contributing to the thermal width depending on the temperature. The peak position and its width
in the spectral function of heavy quarkonium can be translated into
the real and imaginary part of the potential \cite{Rothkopf:2011db}. Other ways to calculate the imaginary potential are presented in \cite{imaginaryot}.

The weakly coupled calculation of the imaginary part of the static potential in a plasma with momentum-space anisotropy and its effects on the thermal widths of the states have been studied in \cite{imaginary1,imaginary2,Dumitru:2009ni,Dumitru:2010id}. It turns out, that all the results and the binding energies depend on
the anisotropic parameter.

There are certain difficulties to study strong coupling
phenomena in QCD quantitatively. On the other hand lattice
simulations are very useful but is still very difficult to use them to study real time phenomena in QGP. An alternative method for studying different aspects of the strongly coupled QGP is the AdS/CFT correspondence \cite{adscft}. Using the gauge/gravity dualities we were able to obtain several important insights into the dynamics of strongly-coupled gauge theories \cite{CasalderreySolana:2011us,DeWolfe:2013cua}. Although the exact gravity dual of the QCD is not known yet it appears to exist a universality among the different gravity dual backgrounds.
An example of one the most known results of AdS/CFT is the prediction of the ratio of shear viscosity over entropy density \cite{etas1}.

Recently some interesting anisotropic models have been developed in \cite{Janik:2008tc,Mateos:2011ix} and \cite{otheraniso}. The
anisotropic  geometry of \cite{Janik:2008tc} has a naked singularity and here we will focus on the gauge/gravity duality \cite{Mateos:2011ix} which is an extension of the solutions of \cite{takayanagi}. The dual theory is a deformed anisotropic finite temperature $\mathcal{N} = 4$ sYM plasma, where the two
directions are rotationally symmetric, while the other one
breaks this symmetry and can be thought to correspond to the beam direction of the plasma.

The static potential in the strongly coupled anisotropic plasma, which is relative to the imaginary potential analysis, was studied in \cite{Giataganas_aniso,Rebhan:2012bw,matqq} where it was found that the static potential is decreased and stronger decrease observed for quark-antiquark pair aligned along the anisotropic direction than the transverse one. The dipole in the plasma wind and the modification of the scaling in screening length for ultrarelativistic velocities outside the transverse plane was found in \cite{matqq}, and the velocity-dependent quark anti-quark separation studied in \cite{Chakraborty:2012dt}. The drag force was studied in \cite{Chernicoff:2012iq,Giataganas_aniso} and jet quenching parameters in \cite{Giataganas_aniso,Chernicoff:2012gu,Rebhan:2012bw}. Moreover, the stopping distance of a light probe at small anisotropy was investigated in \cite{Muller:2012uu}. In \cite{Fadafan:2012qu} the energy loss of a rotating massive quark has been studied and it was argued that the energy loss due to radiation in anisotropic media is less than the isotropic case. All the observables related to the energy loss depend on the relative
orientation between the anisotropic direction and the velocity of
the quark. The
photon production which is an important way to probe the anisotropies in the plasma due to weak interaction of the photons in the later stages of the QGP was studied in \cite{Patino:2012py} and in presence of a constant magnetic field in \cite{Wu:2013qja}. One of the interesting features of
this anisotropic background is that the ratio of the shear viscosity to the entropy density \cite{etas1}  takes non-universal values along the different directions and violates clearly the $1/4 \pi$ bound \cite{Rebhan:2011vd} without the inclusion of any higher
derivative theories in gravity as happened in \cite{Brigante:2008gz}.
Other anisotropic models as in p-wave superfluids, may give non-universal values \cite{Erdmengerab1}. A recent review on the observables on the anisotropic strongly coupled plasmas, can be found in \cite{Giataganas_review}. In this review our results for the imaginary potential were preannounced and discussed briefly.

Motivated by the experimental as well as theoretical studies, we
study the imaginary part of static potential in a strongly coupled
holographic 
anisotropic plasma. We use the top-down geometry
 \cite{Mateos:2011ix} where the anisotropy is generated by a space-dependent axion. In the context of AdS/CFT there are
different approaches which can lead to a complex static potential.
In our case the imaginary potential originates from the fluctuations at the deepest point of the U-shaped string in the
bulk \cite{Noronha:2009da}. Using this method, we find how the imaginary potential depends on the anisotropic parameter. We have also shown that there exist a peculiar turning point in the imaginary part of the potential which is similar to the one of the real part and we elaborate on this point further.

The aim of this work is to analyze the imaginary potential in small and large anisotropies in a holographic anisotropic plasma.
In small anisotropies the pressure inequality along the anisotropic direction and along the transverse direction is similar to the observed QGP, while at larger anisotropies it gets inverted \footnote{This refers to the anisotropy of the QGP, created by the rapid expansion of the plasma along the longitudinal beam axis at the earliest times after the collision. The longitudinal pressure to the beam axis is lower than the transverse one and the momenta of the partons along the beam direction are lower than the ones in the transverse space.}. This gives a naive physical motivation of the study of small anisotropies. Moreover, the entropy at small and large anisotropies scales differently with the temperature, with a transition being around $a/T\simeq 3.7$. In small anisotropies the entropy goes as $T^3$ and in large ones as $T^{8/3}$.
Since the analytic solution of the full background is not known \cite{Mateos:2011ix}, a separate analysis should be performed for large and small anisotropies where the background is known analytically and an analytic calculation of the observables is possible. In the case of the large anisotropies we calculate the observables only numerically. We find that the imaginary potential in the presence of anisotropy for fixed 
temperature decreases, where larger decrease happens along the anisotropic direction. Also the distance that the imaginary potential is generated depends on the anisotropy and is smaller than the isotropic case. Along the anisotropic direction the imaginary potential is generated for smaller distances compared to the ones in transverse direction. To obtain the thermal width we find an analytic form of the real part of the potential using a numerical fitting. We observe that the thermal width is decreased in presence of anisotropy and bigger decrease happens along the transverse plane. When the entropy density is kept fixed
the $ImV$ along the anisotropic direction gets increased in absolute value
while in the transverse direction we observe a decrease.

The paper is organized as follows. In the next section the
imaginary potential for a generic anisotropic background is obtained. In
section 3, we apply the generic formulas to a particular anisotropic
background and analyze the imaginary potential at small and large
anisotropy. Section 4 is devoted to the discussion of alternative approaches for calculating imaginary potential at strong coupling.
In section 5, we discuss our results in comparison with the ones
obtained in the weakly coupled gauge theories. In section 6 we investigate the thermal width of a quarkonium in the strongly coupled
anisotropic plasma. In this section also the real part of
anisotropic potential has been discussed. In the final section we conclude and discuss our results. Our main text is supported by three appendices.\newline
Note added: We have received the paper \cite{Finazzo:2013rqy} where the thermal width and the imaginary part of the static potential are studied in holography, and calculated in the Gauss-Bonnet gravity using a similar method presented here.

 \section{Imaginary Potential in General Background}\label{general-imaginary-potential}
In this section we obtain the imaginary potential for a generic background
which may have anisotropies in the spatial dimensions. The gravity background we use has the following form
\be
 ds^2=G_{00}dx_0^2+G_{11}dx_1^2+G_{22}dx_2^2+G_{33}dx_3^2+G_{uu}du^2,\label{general-background}
\ee
where the metric elements are functions of the radial distance $u$.

We consider the usual orthogonal Wilson loop that corresponds to the
heavy Q\={Q} pair. This can be represented with the static gauge
ansatz for the string and the following radial dependence \be
x_0=\tau,\quad x_p=\sigma,\quad u=u(\sigma)~,\label{static-gauge}
\ee where $x_p$ could be $x_1$, $x_2$ or $x_3$. The analysis of the
action and the resulting energy and the length of the string, useful
also for the static potential, is presented in the Appendix
\ref{app:qq}. Here we use results from this Appendix and for
presentation purposes we define the following quantities
\be\la{fmg1} f(u):=-G_{00} G_{pp}~,\quad g(u):=-G_{00} G_{uu}~. \ee
In order to generate the imaginary part of the potential we need to
consider the fluctuations around the $u_0$ of the string worldsheet
that has a turning point $u_0$ close to $u_h$. These fluctuations will be
responsible for the complex part of the potential. Fluctuations of
the form $u(\s)\rightarrow u(\s)+\delta u(\s)$ such that $\delta
u'\rightarrow 0$ result to the following string partition function
\be \mathcal{Z}_{\text{string}}\sim\int\mathcal{D}\delta
U(\sigma)~e^{i~S_{NG}(u+\delta u)}\sim \int d(\delta U_{-N})\dots
d(\delta U_N)~e^{-i \frac{\mathcal{T}
\Delta\sigma}{2\pi\alpha'}\sum_j\sqrt{f+g~u'^2_{j}}}, \ee where the
integral is divided to $2N$ parts such that $x_N=L/2$ and
$x_{-N}=-L/2$, $\mathcal{T}$ is the long time edge of the Wilson
loop and the action \eq{sgener1} is used. The profile of this
configuration is a string worldsheet which exhibits a turning point
$u_0$ at $\s=0$. Expansion around this point considering the
fluctuations leads to \bea\nn &&u(\s)\simeq u_0+\s^2 u_0''/2,\qquad
f(u)+g(u) u'^2 \simeq c_2+ c_1 \s^2\\\nn
&&c_1\simeq \frac{1}{2}u_0''~\left[2(g_0+\delta u ~g_0') ~u_0''+f_0'+\delta u~ f_0''\right]\simeq\frac{1}{2}u_0''~\left(2u_0''~g_0+f_0'\right)=\frac{{f'_0}^2}{2 g_0}\\
&&c_2\simeq f_0+\delta u~{f'_0}+\frac{1}{2}\delta u^2 f_0''
\eea
where $f_0:=f(u_0),~g_0:=g(u_0)$ and the derivatives $f_0':=\partial_u f(u)|_{u=u_0}$. To obtain $c_1$ we have used the variation of equation of motion \eqref{uprime01} which gives $u_0''=f_0'/2 g_0$. To generate the imaginary part of the potential, $c_2$ needs to be always negative and more particularly $\delta u_{1,j}<\delta u_j<\delta u_{2,j}$ where $\delta u_{1,j}$ and $\delta u_{2,j}$ are the roots of the $c_1\s_j^2+c_2$, so the relevant quantity under the square root becomes negative.

Using the saddle-point approximation\footnote{Note that its accuracy depends
on the particular background used.} it is straightforward to show
that $\delta u=-{f_0'}/{f_0''}$ and therefore
\be
c_1\,\s^2+c_2 \simeq \frac{{f'_0}^2}{2
g_0}\,\s^2+f_0-\frac{{f'_0}^2}{2f''_0}~. \label{C1-C2}
\ee
The real and imaginary parts of the potential are separated  in the partition function as
\be
 \exp[-i~\mathcal{T}~V_{Q\bar{Q}}]
 \sim\exp\bigg[-\frac{\mathcal{T}}{2\pi\alpha'}\bigg(\int_{|\s|<\s_c}d\s \sqrt{-c_2-\s^2 c_1}
+i\int_{|\s|>\s_c}d\s \sqrt{f(u)+g(u)~{u'}^2}\bigg)\bigg]\label{VQQ}
\ee
where
\bea\nn
&&\s_c=\sqrt{-c_2/c_1}\quad if\quad c_2<0\\
&&\s_c=0\quad if\quad c_2 >0~.\la{c2c1}
\eea
To have imaginary part we need $\s_c\neq 0$, which activate
the worldsheet fluctuations around $u_0$ and close to $u_h$.
The complex part comes from the first term in the partition
function and using \eqref{C1-C2} takes the following compact form
\be
\text{Im} V_{Q\bar{Q}}=
\frac{1}{4\alpha'}\frac{c_2}{\sqrt{c_1}}=
\frac{1}{2\sqrt{2}\alpha'}\left[\frac{f_0}{|f_0'|}-\frac{|f_0'|}{2f_0''}
\right]~\sqrt{g_0}~.\label{im-potential2}
\ee
This is a generic formula that gives the imaginary part of the potential in terms of the metric elements of a background \eq{general-background}, \eq{fmg1}. $\text{Im}V$ can be expressed in terms of the length $L$  of the Wilson loop instead of $u_0$ using the equation \eq{staticL1}.
For the special case of $g\prt{u}=1$ we recover the results of \cite{Noronha:2009da}. Note that for very large distances there is a possibility of additional configurations contributing to our result according to \cite{Bak:2007fk}, but here we are interested mostly on distances $L T<1$.

The $\text{Im}V$  shifts the Bohr energy level $E_0\rightarrow E_0 +\delta E -
i\Gamma$ where $\Gamma$ is given by
\be
\Gamma=-<\psi|\text{ImV}_{Q\bar{Q}}|\psi>,\label{gamma}
\ee
with $|\psi>$ being the ground state of the unperturbed static potential,
which in $\cN=4$ sYM it is the usual Coulomb potential. In the
following we apply the generic formulas derived here to a particular
anisotropic background.

\section{Imaginary potential in anisotropic background}
In this section we apply the results from section
\ref{general-imaginary-potential} to find the imaginary part of the
potential between quark anti-quark in a spatially anisotropic
background. We give a short review on the axion space dependent background
 introduced in \cite{Mateos:2011ix}. The anisotropic
background is known analytically for small anisotropy and numerically in the
opposite limit.

\subsection{Anisotropic Background}

The anisotropic background \cite{Mateos:2011ix}
comes from a deformation of $\cN=4$ SYM in finite temperature. From
the gravity side the deformation can be thought as generated from a
backreacting number of $D7$ branes wrapping the internal sphere and
two of the spatial directions in the spacetime, therefore deforming
the space anisotropically. The gravity background includes an
axion, sourced by the D7 branes, depending on the single spatial
anisotropic direction, which in the dual field theory is translated
to a $\theta$-term depending on the same direction.

Using the radial coordinate $u\in[u_h,\infty)$ the background in
string frame is given by\footnote{We set the radius of the AdS part
equals to one for simplicity and we recover it once we need it.}
\bea &&ds^2 =
 u^2\left( -\cF \cB\, dx_0^2+dx_1^2+dx_2^2+\cH dx_3^2\right) +\frac{ du^2}{u^2 \cF}+ e^\frac{\phi}{2} \, d\Omega^2_{S^5}\,.
  \label{aniso-bg} \\
&&  \chi = a x_3, \qquad \phi=\phi(u) \,,\qquad 
\label{chi}
\eea
where $a$ is the anisotropy parameter, $\cF, \cB$ and $\cH$ are functions of radial coordinate $u$. The temperature and the entropy density of the solution are respectively given by
\bea
T=\frac{u_h^2|\cF'_h|\,\sqrt{\cB_h}}{4\pi},\hspace{20mm}s=N_c^2\,\frac{u_h^3}{2\,\pi}\,e^{-\frac{5}{4}\phi_h},
\eea
where we have used $G_5=\pi/2N_c^2$ and $N_c$ is the number of colors.

In the range of small $a/T$ the analytic solution is of the form
\bea
\cF(u) &=& 1 - \frac{u_h^4}{u^4} + a^2 \cF_2 (u)  +\mathcal{O}(a^4)\nn\\
\cB(u) &=& 1 + a^2 \cB_2 (u) +\mathcal{O}(a^4)\,, \label{small-a-T}\\
\cH(u)&=&e^{-\phi(u)},\quad\mbox{where}\quad \phi(u) =  a^2 \phi_2 (u)  +\mathcal{O}(a^4)\,,\nn
\eea
in second order, where the functions appear are
\bea
&&\mathcal{F}_2=\frac{u_h^2}{24 u^4}\left[\frac{8(u^2-u_h^2)}{u_h^2}-10\log2+\frac{3 u^4+7 u_h^4}{ u_h^4}\log\left(1+\frac{u_h^2}{u^2}\right)\right],\\
&&\mathcal{B}_2=-\frac{1}{24\,u_h^2}\left[\frac{10 u_h^2}{u^2+u_h^2}+\log\left(1+\frac{u_h^2}{u^2}\right)\right],\\
&&\varphi_2=-\frac{1}{4\,u_h^2}\log\left(1+\frac{u_h^2}{u^2}\right).
\eea
The temperature and the entropy density to order $a^2$ are respectively given by
\be
T=\frac{u_h}{\pi}+\frac{5\log 2-2}{48\,\pi\,u_h}\,a^2+\mathcal{O}(a^4),\hspace{17mm}
s=\frac{\pi^2 N_c^2 T^3}{2}+\frac{N_c^2 T}{16} a^2+\mathcal{O}(a^4).\label{temp-aniso}
\ee
The pressures and energies which are useful on making a naive
connection with the weak coupling results can be found from the
expectation values of the stress energy tensor and read\footnote{We
 use the notation $\parallel$ for the anisotropic direction $x_3$, and
$\perp$ for transverse directions $x_1$ and $x_2$.}
\bea
\label{ppxx}
&&E= 3 e_1+ a^2 e_2+{O}(a^4)~,\quad P_{\perp}= e_1+ a^2
e_2+{O}(a^4)~,\quad P_{\parallel}= e_1- a^2 e_2+{O}(a^4)~,\\\nn
&&\mbox{where}\quad e_1=\frac{\pi^2 N_c^2  T^4}{8}~,\quad
e_2=\frac{ N_c^2  T^2}{32}~. \eea
Notice that for the pressures at small anisotropy we have the following inequality
\be\label{ppp1} P_\parallel<P_\perp~. \ee

\subsection{Analysis of the Imaginary Potential}

To study the Q\={Q} potential we consider the dipole on the $x_1x_3$
plane at an angle $\theta$ with the $x_3$ axis. This can be done by
making for example a generic transformation
\bea
&&x_1\rightarrow x_1\sin\theta+x_3\cos\theta~,\nn\\
&&x_3\rightarrow x_1\cos\theta-x_3\sin\theta~ \eea which corresponds
to a rotation of the $x_1x_3$ plane. We consider the new $x_1$
coordinate in the static gauge ansatz as in Appendix \ref{app:qq},
where the constant angle $\theta$ will direct the dipole in the
$x_1x_3$ plane with $x_1=x_{3,"original"}:=x_\parallel$ for $\th=0$
and therefore the dipole aligned along the anisotropic direction, and
$x_1=x_{1,"original"}:=x_\perp$ for $\th=\pi/2$ where the dipole is
aligned along the transverse plane to anisotropic direction. Any
other angle between $\prt{0,\pi/2}$ rotates the dipole from the
anisotropic direction to the transverse plane. In this case the
defined functions \eq{fmg1} take the form \bea
&&f=-G_{x_0x_0} G_{x_1 x_1}=u^4\cF \cB(\sin^2\theta+\cH\cos^2\theta),\nn\\
&&g=-G_{x_0x_0} G_{uu}=\cB.\label{VW-aniso}
\eea

\subsubsection{Imaginary Potential in small Anisotropy}\label{small-aniso}
The small $a/T$ limit of the imaginary potential needs to be studied separately from the large limit, for several reasons analyzed in the introduction. In this section, we study the imaginary potential in small anisotropy, where the pressure inequality \eq{ppp1} is the desired one, observed in the anisotropic QGP, and the background is known analytically. For the small anisotropy one can find the explicit form of the imaginary potential to order $a^2$, using the previous derived equation \eq{im-potential2}
\bea\nn
\text{Im}V_{Q\bar{Q}}&=&\frac{1}{24\,\sqrt{2}\,\alpha'}\cdot
\bigg\{\frac{1-3\,\tilde{u}_0^4}{\tilde{u}_0}u_h+a^2 \frac{1}{144\, u_h\, (1+\tilde{u}_0^2)^2\tilde{u}_0}\,\bigg[\tilde{u}_0^2\bigg(-15 -9\,\tilde{u}_0^2 + 194 \,\tilde{u}_0^4  \\
&&+ 130 \,\tilde{u}_0^6 - 63 \,\tilde{u}_0^8 - 45 \,\tilde{u}_0^{10} +3 \,(1 + \tilde{u}_0^2) (5-4\,\tilde{u}_0^2-26\,\tilde{u}_0^4+ 9\,\tilde{u}_0^8) \cos2\theta\nn\\
&&-180\tilde{u}_0^2(1+\tilde{u}_0^2)^2 \log2\bigg)+3(1+\tilde{u}_0^2)^2(63\tilde{u}_0^4-1)
\log\left(1+\tilde{u}_0^2\right)\bigg]\bigg\}+\mathcal{O}(a^4).\label{im-aniso1}
\eea
where $\tilde{u}_0=u_h/u_0$.
We are interested in expressing the imaginary potential as a function of $L$, $T$, $a$ and $\theta$ which are the distance between $Q\bar{Q}$, temperature of the anisotropic plasma, the anisotropy parameter and the angle of dipole to the anisotropy direction respectively. In general this is not doable analytically since the equation \eqref{staticL1} can not be integrated and solved for $u_0$ always. However in the region of small $LT$ the results can be fitted numerically and an approach is presented in Appendix \ref{app:imsml}. The  imaginary potential in this region is given by \eq{im-aniso1} and is of the form
\be
\frac{1}{T}\,\text{Im}V_{Q\bar{Q}}=V_0(LT)+\frac{a^2}{T^2}\,V_1(LT,\th)+\mathcal{O}(a^4)~,
\ee
where $V_0$ is the zero anisotropy result and the $V_1$
is a complicated function of $LT$, and $\th$ and the $a^2/T^2$ term leads to increase of the absolute value of the potential in any direction.

The problem can also be solved in any region of $L$ numerically. The critical bulk distance of the world-sheet where the potential gets a non-zero imaginary part for a particular critical length $L$, can be found by setting $c_2$ of equation \eq{c2c1} equal to zero  and solving it for $u_0$. The corresponding critical length then can be derived by the integration of \eq{staticL1}. The imaginary potential in terms of length $L$  can be subsequently found from \eq{im-potential2}.

\begin{figure*}[!ht]
\begin{minipage}[ht]{0.5\textwidth}
\begin{flushleft}
\centerline{\includegraphics[width=70mm]{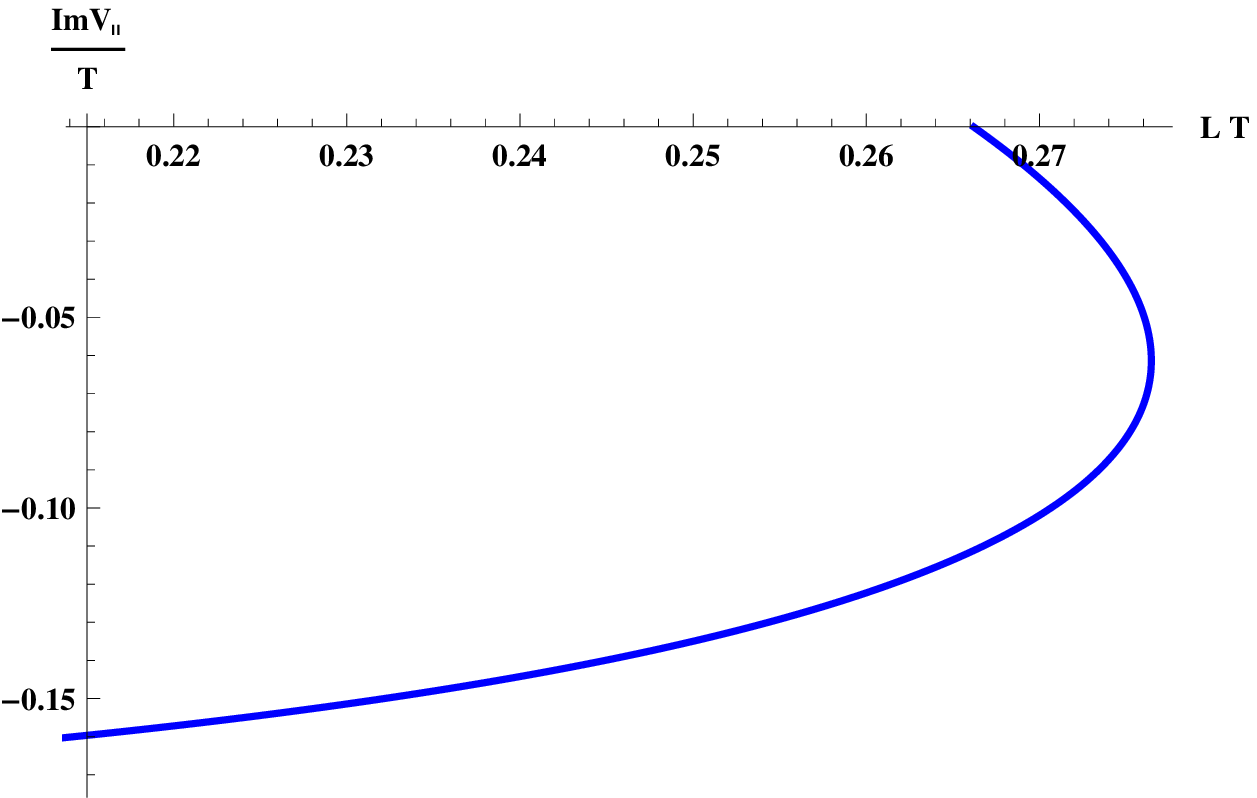}}
\caption{\small{The imaginary potential $ImV_\parallel$ for $a/T=0.3$ in anisotropic direction. The imaginary part is generated for $L_{0,\parallel}\simeq 0.266$ and decreases until length  $L_{m,\parallel}\simeq 0.277$. The shape of the curve of the imaginary potential along the transverse direction, or for the isotropic case are the same without any crossing between the curves. Our method generating the imaginary potential gives a turning point at $L_m$ and therefore we are interested only on the upper branch of the solution.
\vspace{0cm}}}\label{fig:l1}
\end{flushleft}
\end{minipage}
\hspace{0.3cm}
\begin{minipage}{0.5\textwidth}
\begin{flushleft}
\centerline{\includegraphics[width=70mm]{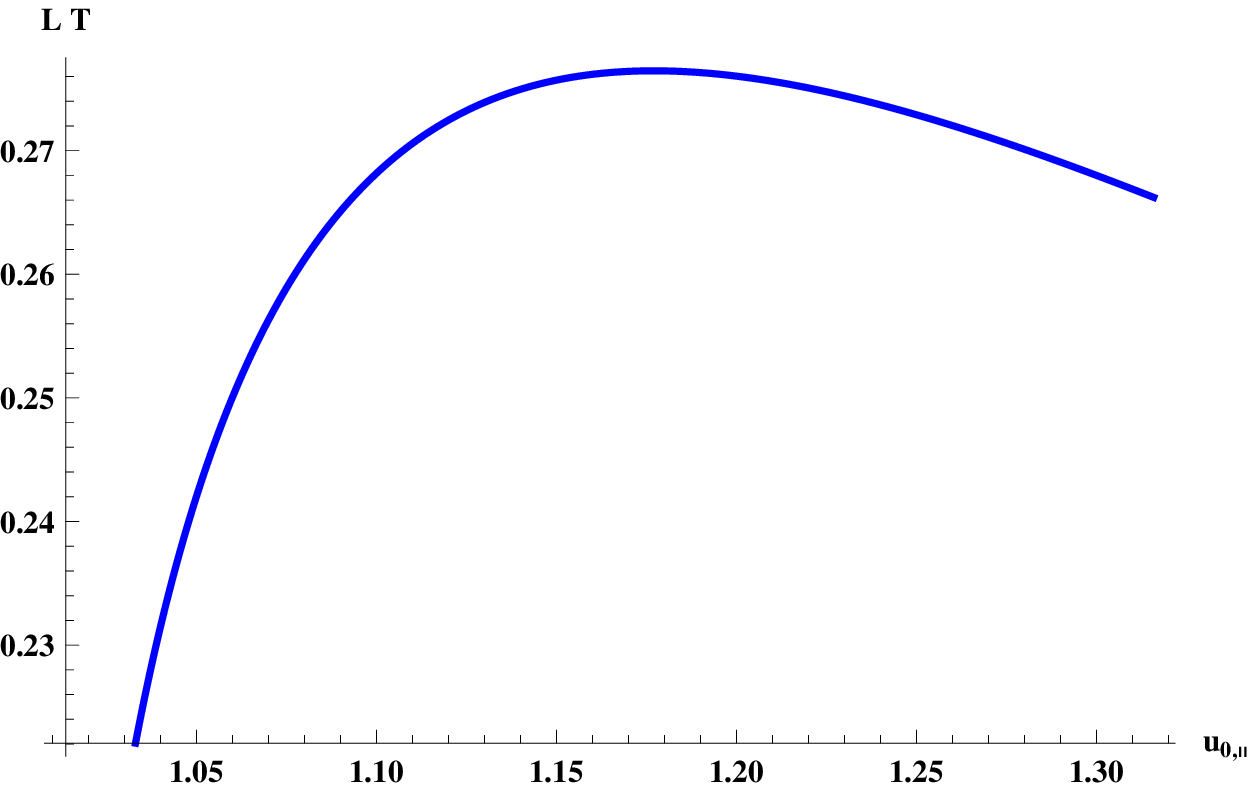}}
\caption{\small{The length $L$ in terms of the radial distance $u$. The turning point corresponds to the $L_{m,\parallel}$. The region we study is the right branch to the $L_m$ which generate the upper branch of the solution in \ref{fig:l1}.
\vspace{1.8cm}}}
\label{fig:l2}
\end{flushleft}
\end{minipage}
\end{figure*}

The imaginary part in the anisotropic background starts to be generated in smaller distances than in absence of anisotropy
\be\la{lrel}
L_{0,\parallel}<L_{0,\perp}<L_{0,~iso}~.
\ee
When the imaginary potential is created, it is increased in absolute value with $L$ until a value $L_m$ or $u_m$ where $\partial_u L\prt{u}$ is zero as shown in Figures \ref{fig:l1} and \ref{fig:l2}.
After this value the imaginary potential develops a second solution that is not acceptable due to its boundary behavior. Therefore we study the region $\prt{L_0,~L_m}$. For the critical distances $L_m$ holds the same  inequality relation \eq{lrel} for the different directions.
As an example for $a/T=0.3$, the critical length for the two directions and the isotropic background case are
\bea
L_{0,\perp}T\simeq0.26626~, L_{0,\parallel}T\simeq0.26618~, L_{0,~iso}T\simeq 0.26634~
\eea
and for the $L_m$
\bea
 L_{m,\perp}T\simeq0.27654~, L_{m,\parallel}T\simeq0.27646~,L_{m,~iso}T\simeq 0.27664~.
\eea
The differences due to anisotropy in this example are small because we are in  the small $a/T$ limit, where the background is deformed compared to the isotropic only with $\prt{a/T}^2$ terms. In the next section we show that for larger anisotropies the equation \eq{lrel} remains similar.

\begin{figure*}[!ht]
\begin{minipage}[ht]{0.5\textwidth}
\begin{flushleft}
\centerline{\includegraphics[width=70mm]{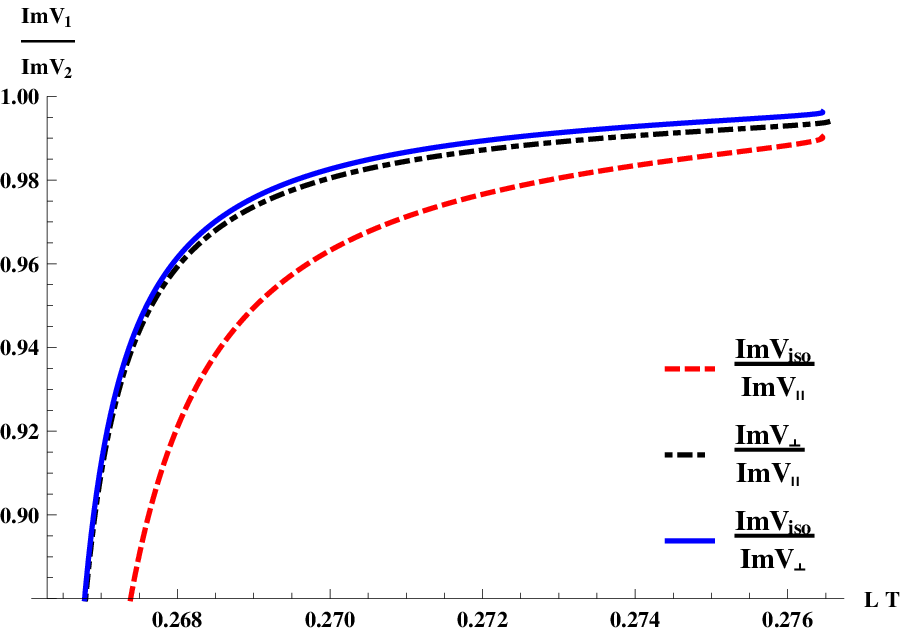}}
\caption{\small{The ratios of the imaginary potentials along the different directions with anisotropy $a/T=0.3$ with respect to distance $L$. From the plot is clear that $|ImV_{iso}|<|ImV_{\perp}|$$<|ImV_{\parallel}|$ . Settings: $ImV_{iso}/ImV_{\perp}$: blue solid, $ImV_{iso}/ImV_{\parallel}$: red dashed, $ImV_{\perp}/ImV_{\parallel}$: black dot-dashed.
\vspace{0cm}}}\label{fig:im1}
\end{flushleft}
\end{minipage}
\hspace{0.3cm}
\begin{minipage}{0.5\textwidth}
\begin{flushleft}
\centerline{\includegraphics[width=70mm]{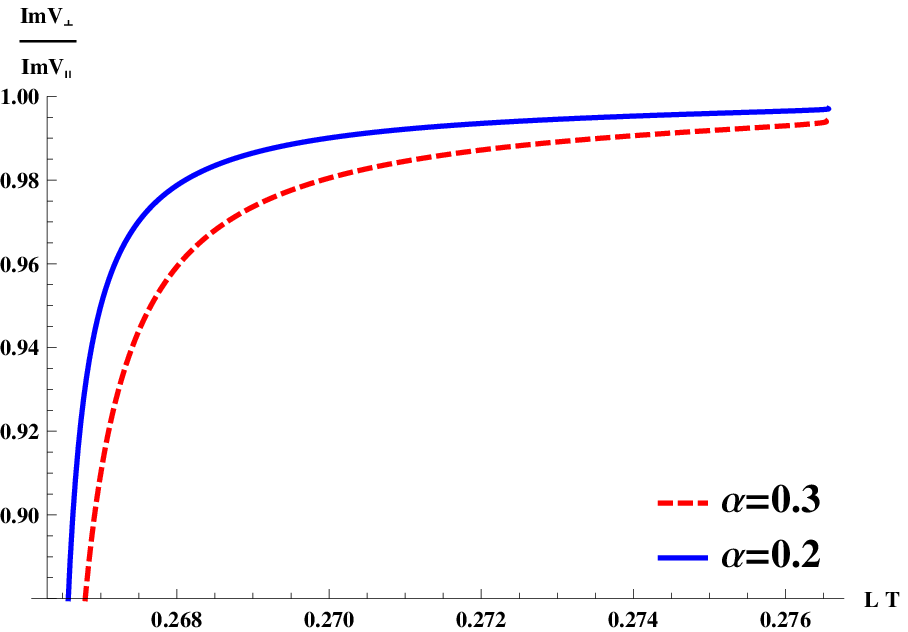}}
\caption{\small{The ratio $ImV_{\perp}/ImV_{\parallel}$ depending on the distance $L$ for two different anisotropic parameters $a/T=0.3,$ and $0.2,~\prt{T=1}.$ We find that increase of the anisotropy leads to increase of the difference of the values of anisotropic potentials along the different directions.  \vspace{0.5cm}}}
\label{fig:im2}
\end{flushleft}
\end{minipage}
\end{figure*}

In Figure \ref{fig:im1}, we show how the anisotropy modifies the imaginary potential along the different directions. The ratios of $ImV_{iso}/ImV_{\perp}$, $ImV_{iso}/ImV_{\parallel}$ and $ImV_{\perp}/ImV_{\parallel}$ are plotted and are all smaller than the unit. We find that in absolute value the imaginary potential is increased due to anisotropy with larger increase along the anisotropic direction
\be\la{imin}
|ImV_{iso}|<|ImV_{\perp}|<|ImV_{\parallel}|~,
\ee
as shown in Figure \ref{fig:im1}.
Increase of the anisotropy leads to further increase of the imaginary parts respecting the inequality \eq{imin} in the region of small $a/T$.
Moreover the ratios have large deviations around the critical distance where the potentials take smaller values and this is the reason of the sudden decrease of the curves in the plots.

\begin{figure*}[!ht]
\begin{minipage}[ht]{0.5\textwidth}
\begin{flushleft}
\centerline{\includegraphics[width=70mm]{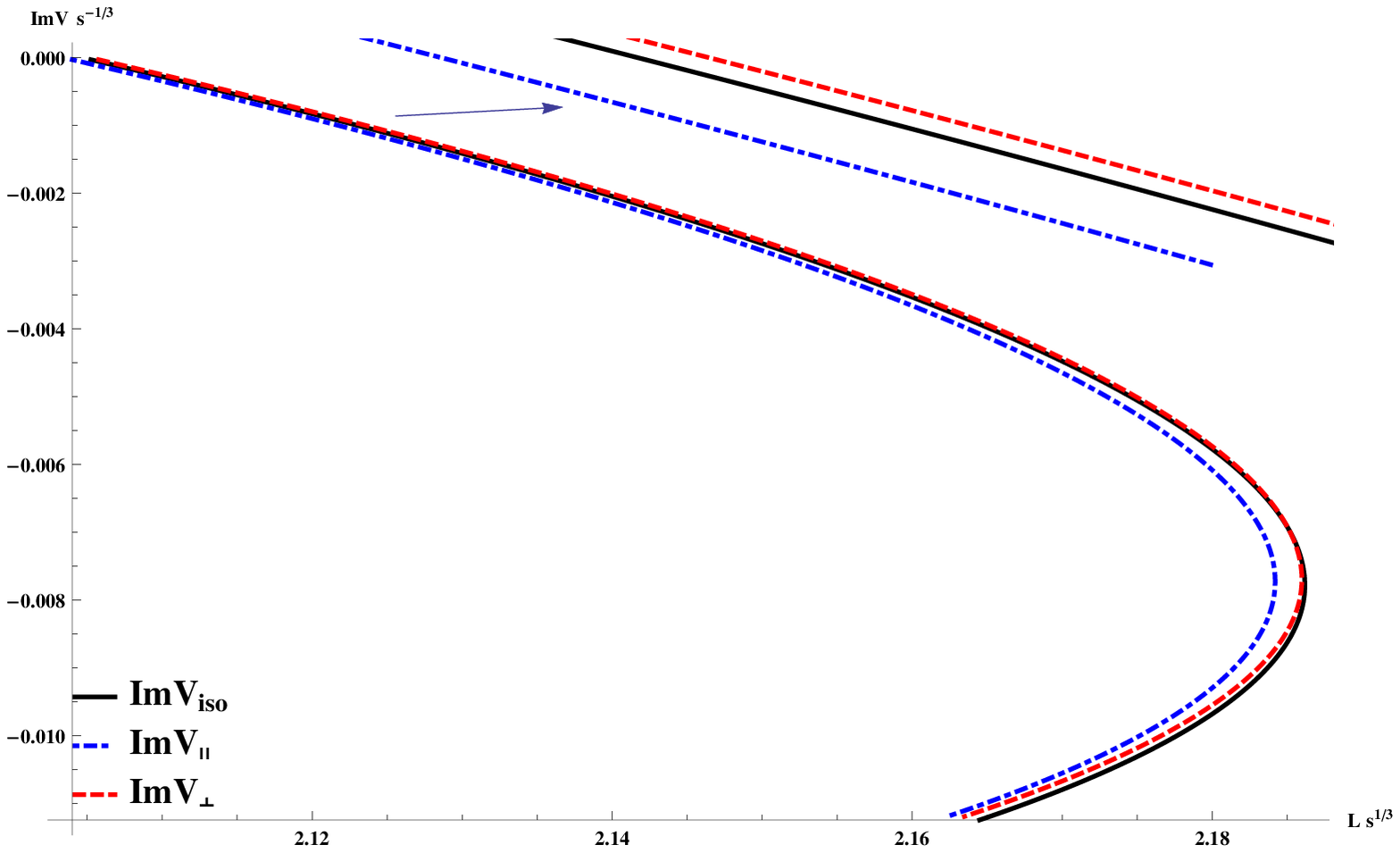}}
\caption{\small{The $ImV$ with fixed entropy density with as a function of length $L$. The zoomed area shows clearly the pattern.  Settings: $ImV_{iso}$: black solid, $ImV_{\parallel}$: blue dot-dashed, $ImV_{\perp}$: red dashed.
\vspace{0cm}}}\label{fig:ime1}
\end{flushleft}
\end{minipage}
\hspace{0.3cm}
\begin{minipage}{0.5\textwidth}
\begin{flushleft}
\centerline{\includegraphics[width=70mm]{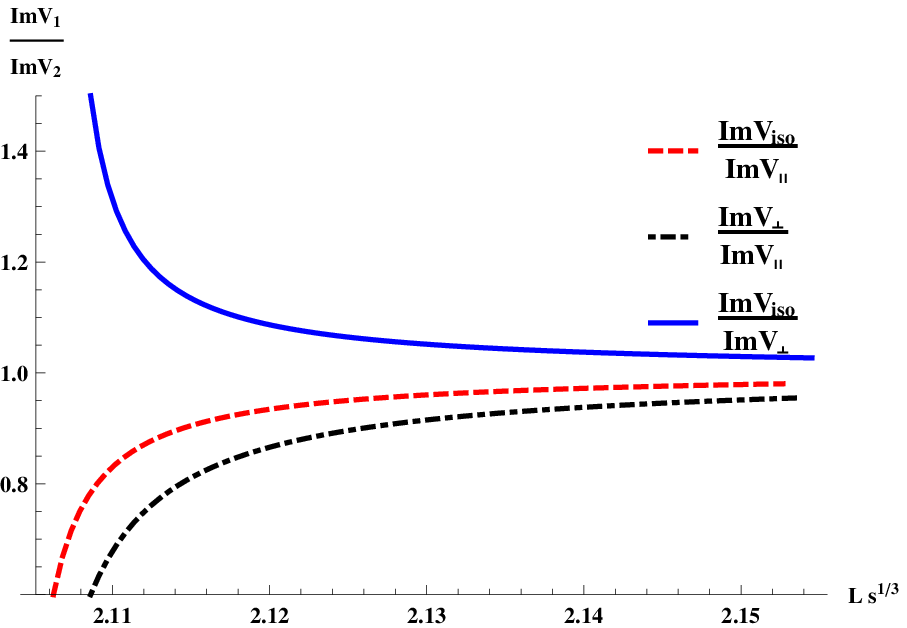}}
\caption{\small{The ratios of the imaginary potentials along the different directions with anisotropy with respect to distance $L$ for fixed entropy density. From the plot it is clear that $|ImV_{\perp}|<|ImV_{iso}|$$<|ImV_{\parallel}|$ . Settings: as in Figure \ref{fig:im1}.  \vspace{0.5cm}}}
\label{fig:ime2}
\end{flushleft}
\end{minipage}
\end{figure*}

Concerning the parallel and transverse directions to anisotropic we find that increase of the anisotropy leads to increasing difference in the values of the potential along the two different directions as can be seen in Figure \ref{fig:im2}.
Fixing the entropy density we find that the $ImV$ along the anisotropic direction gets increased in absolute value while in the transverse direction we observe a decrease. The results are presented in Figures \ref{fig:ime1} and \ref{fig:ime2}.
Summarizing some of the results of this section we find that  in
presence of anisotropy, when the temperature
is kept fixed, the imaginary potential is increased in
absolute value with the largest increase being along the anisotropic
direction. Increase of the anisotropy leads to further increase of
the imaginary parts, with the one in the anisotropic direction
increasing more rapidly. This is something  expected since the
metric in this direction is more heavily dependent on the parameter
$a$. We also find that the distance that the imaginary potential is
generated depends on the anisotropy, and it takes non zero value for
smaller lengths in presence of anisotropy. Along the anisotropic
direction the imaginary potential is generated for smaller distance
than in the transverse direction. When the entropy density is kept fixed
the $ImV$ along the anisotropic direction gets increased in absolute value
while in the transverse direction we observe a decrease.

We will apply these results in the section \ref{Thermal-width}  to
find the thermal width for a heavy quarkonia. In next subsection for
completeness we study the imaginary potential for large anisotropy.

\subsubsection{Imaginary Potential in large
Anisotropy}\label{large-aniso}

For completeness we study the large anisotropy region where the
solutions in our background are not known analytically, but numerical
solutions are possible \cite{Mateos:2011ix}. Estimations
for the parameter $a/T$ suggest that a value which would correspond
to the observed anisotropic QGP would be at least $a/T\gtrsim 4$
\cite{Giataganas_aniso}. In this region however the pressure
inequality \eq{ppp1} is inverted and it is not anymore clear the
direct relation to the QGP.

\begin{figure*}[!ht]
\begin{minipage}[ht]{0.5\textwidth}
\begin{flushleft}
\centerline{\includegraphics[width=3in]{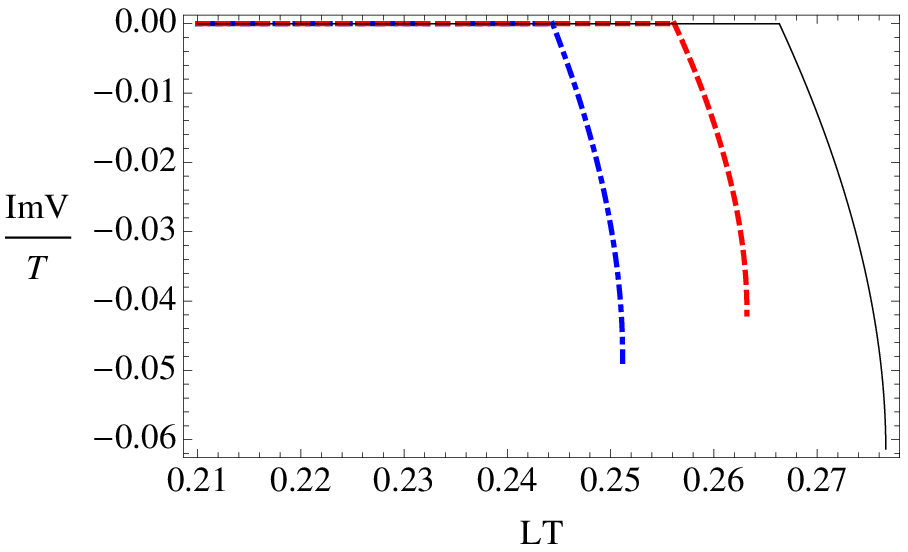}}
\caption{\small The imaginary potential for $a/T\simeq4.4$ versus LT. Settings: as in Figure \ref{fig:ime1}. \vspace{0.3cm}}\label{large-ImVa}
\end{flushleft}
\end{minipage}
\hspace{0.1cm}
\begin{minipage}[ht]{0.5\textwidth}
\begin{flushleft}
\centerline{\includegraphics[width=3.2in]{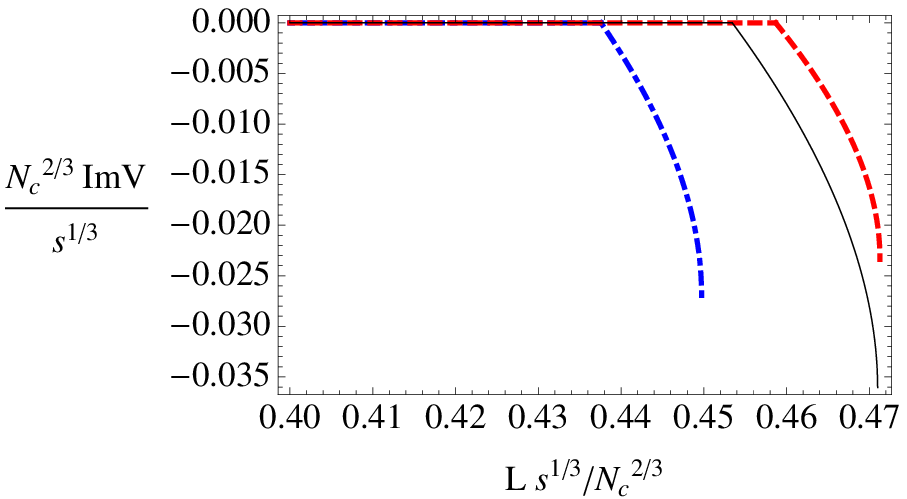}}
\caption{\small The imaginary potential for $a N_c^{2/3}/s^{1/3}\simeq2.465$ versus $L s^{1/3}/N^{2/3}$. Settings: as in Figure \ref{fig:ime1}.}\label{large-ImV-enta}
\end{flushleft}
\end{minipage}
\end{figure*}

The absence of a full analytical background solution forces us to
work numerically solving the supergravity equations for the metric with boundary at infinity \eq{aniso-bg}. We check the accuracy of saddle-point
approximation for the string solution by considering the Nambu-Goto action
in the anisotropic background with fixed $a/T$.
We find that for relatively low anisotropies $a/T \lesssim 12$ the approximation
is better. By using the equations \eq{im-potential2} and
\eq{staticL1} we obtain the relation
$\text{ImV}_{Q\bar{Q}}=\text{ImV}_{Q\bar{Q}}(L, T, a, \theta)$ and our findings are
plotted in Figures \ref{large-ImVa}. Notice the increase of the distance between the
curves for the different directions and to the isotropic results
comparing to the previous plots for small anisotropies. The
behavior of imaginary potential is similar to the small anisotropy
case $L_{0,\parallel}<L_{0,\perp}<L_{0,~iso}$.

Using our results for imaginary potential we find the thermal width
in section \ref{Thermal-width} for small and large anisotropy and we
compare the results with isotropic case.

\section{Alternative Imaginary Potential Approaches in Strong Coupling}

In our approach the imaginary potential originates from the
fluctuations at the deepest point of the string in the bulk. In the
context of AdS/CFT there are several other approaches which can lead
to a complex static potential, for example \cite{0807.4747} and \cite{1211.4942}. In this section we elaborate further on the peculiar turning point $L_m$ appearing in the imaginary part of the potential and
we discuss ways of avoiding its appearance with alternative approaches.
A straightforward approach, is to consider extended range of the radial distance $u$
in such a way that the string world-sheet solutions of the Wilson loop corresponding to the static potential become
complex, making the static potential to develop an
imaginary part.

This has been done in finite temperature $\cN=4$ sYM Wilson loop  \cite{0807.4747}. The  string world-sheet has a catenary-like shape where at some critical radial distance $u_c$ the energetic favorable configuration turns out to be two disconnected straight strings. This corresponds to a "screening length". There is a second critical distance at which the solutions to the equation of motions develop an imaginary part $u_{im}>u_c$, due to the form of the hypergeometric functions appear in the solutions of \cite{Rey}. The authors of \cite{0807.4747} choose to introduce a modified normalization scheme for the subtraction of the divergences, subtracting only the real contributions for very large Wilson loops, and making sure that this is consistent with the zero temperature solution.
In their approach the imaginary part of the potential is generated for $u>u_{im}$ and is an absolute value increasing function with respect to length of the Wilson Loop $L$. Increase of temperature makes the imaginary potential to growth faster and therefore leads to stronger associated decay rate.

\begin{figure}
\centerline{\includegraphics[width=75mm]{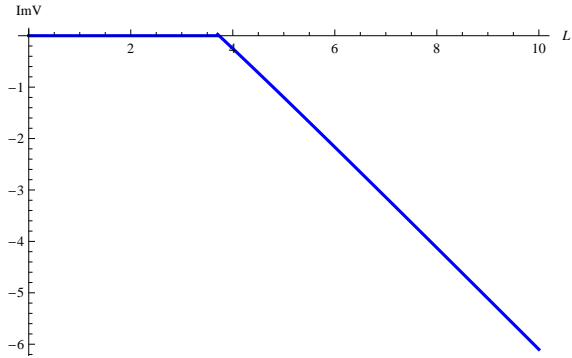}}
\caption{\small The shape of the Imaginary potential using a modified subtraction scheme for the divergences. }\label{fig:alternative}
\end{figure}

The qualitative picture of the imaginary potential is shown in Figure \ref{fig:alternative}. For larger Wilson loops lengths $L_{im}$ which
correspond to $u_{im}$, the imaginary part of the static potential deviates from zero with almost a linear form, leading to a decay of the Q\={Q}.

For completeness it should be also mentioned that the real part of the potential in this approach develops an unexpected $1/L^4$ behavior,
$Re[V(r)]\propto L_s^3/\prt{L\prt{L_s+L}^3}$,
where $L_s\simeq 2.7 u_h$, and is obviously different than the usual ones obtained so far. Moreover, the modified way of subtracting the UV divergences, is not equivalent to substraction of the energy of the two infinite strings at finite temperature which is equivalent to the Legendre transform method on the minimal surface. Nevertheless, it would be interesting to use this alternative approach to compare with our results.

\section{Imaginary Potential in Weakly Coupled Plasma}

It is interesting to attempt a naive comparison of our results with
the ones obtained in weakly coupled theories. In this attempt we
keep in mind that the anisotropies in strong and weak coupling are
generated in different ways, and flavor degrees of freedom are not
considered in our model. On the other hand, although anisotropies
are generated in different ways they can be naively related through
the pressure inequalities, and it is possible that the modifications
on the physical observables depend on the anisotropy and pressure
inequality itself and not on the way that is generated. Therefore a
naive comparison and discussion of the weak coupling results it is
well motivated.

To initiate the connection and the discussion of the weakly coupled
plasmas we are thinking along the lines of \cite{Giataganas_aniso}.
The kinematic example of the collision is that we consider the
accelerated beams of nucleons to move along the $x_3$ anisotropic
direction. The plasma that is created after the collision can be
though as having a  distribution function
$f(t,\textbf{x},\textbf{p})$ which is homogeneous in position space,
anisotropic in momentum space and can be written as
\cite{anisofunction} \be\label{faniso}
f_{aniso}(\textbf{p})=c_{norm}(\xi) f_{iso}(\sqrt{\textbf{p}^2+\xi
(\textbf{p}\cdot \textbf{n})^2})~, \ee where the vector
$\textbf{n}=(0,0,1)$ is the unit vector along the anisotropic
direction and the parameter $\xi$ is the anisotropic parameter. For
$\xi=0$ we are in the isotropic case while for $\xi>0$ the
distribution is contracted in the anisotropic direction and it
corresponds to the plasma created after the heavy ion collisions.
This is the region we focus here.

In Appendix \ref{we-soa} we show that the range of $T\gg a$ we can
relate the weak and strong coupling anisotropic parameters of our
model as \be\label{xiaa} \xi\backsimeq \frac{5 a^2}{8 \pi^2 T^2}~.
\ee In \cite{imaginary1} the authors have used the hard thermal loop
effective field theory to calculate the imaginary potential by
calculating the leading viscus corrections to the gauge-boson self
energies and resumed propagators. In the weakly coupled plasma
$f_{iso}$ can be taken as thermal ideal gas Bose or Fermi
distribution respectively. By calculating the gluon propagator it
can be found the heavy static potential due to one gluon exchange.
The imaginary part of the potential comes from the Fourier transform
of the symmetric gluon propagator. It has been found that the effect
of the anisotropy causes decrease in the absolute value of the imaginary potential,
or equivalently increase it in the negative axis compared to the
isotropic case. Larger increase happens in the transverse to the
anisotropic direction and it becomes more noticeable for larger
separation lengths of the Q\={Q} pairs. This is different to our results where at fixed temperature we find increase of the absolute value of the imaginary potential and
keeping fixed the entropy density we find increase along the anisotropic direction and decrease along the transverse one.
The estimation of the decay
width in this model gives a significant decrease in presence of
anisotropy. For the realistic values of $\xi\simeq 1$ the decrease
is equal to the half of the isotropic equilibrated plasma. In
\cite{1212.2803} the analysis lead to similar qualitative results by
considering an additional dielectric function from the hard loop
resummed gluon propagator.

The behavior of the static potential remain the same in case that soft and hard gluons are seperated. In \cite{0903.3467} the soft gluons are in equilibrium while the hard gluons out of equilibrium. There similarly the Hard thermal loop gluon self energy is calculated with a distribution similar to \eq{faniso} and the appropriate components are taken in the static limit to obtain the static potential. The imaginary potential appears to be independent of anisotropy when the temperature of the non-equilibrium system is obtained isentropically from the equilibrium one. However when the non-equilibrium state is taken with less entropy of the initial one then the imaginary potential is decreased in absolute value and for $\xi\simeq 1$ the dissociation temperature is increased by ten percent compared to the isotropic case.

Moreover when the normalization of \eq{faniso} changes in a way to  preserve the particle number in an ideal gas the effects of anisotropy are much weaker to the static potential \cite{0908.1746}. Even at high anisotropies the quarkonium related observables have very close values to the isotropic results. It seems that
a significant factor for the enhanced results due to the anisotropic distribution function, is the change of the number density of particles with large momentum along the anisotropic direction as indicated by \eq{faniso}, when the normalization factor is kept constant.

\section{Thermal Width of quarkonia in a hot Anisotropic Plasma}\label{Thermal-width}

In order to calculate the thermal width we need a semi-analytic formula of the real part of the potential which will specify the shifted part of the Bohr energy level.

\subsection{Real part of the potential}\label{real-potential}

In this section we attempt to extract an analytic form of the real part of the potential building on the findings of \cite{Giataganas_aniso}. The basic generic formulas of this section are given in the appendix \ref{app:qq}.

To find the static potential in the desirable form $V(L)$, the equation \eq{staticL1} needs to be integrated analytically, then solved as
$u_0(L)$ and inserted in the analytically integrated equation \eq{staticE1}. However, usually either the integrations are not doable analytically or the inversion $u_0(L)$. This is the case for the anisotropic theory under examination. What we  do here is to obtain the analytic form of the static potential fitting the numerical results to a formula derived with dimensional analysis arguments and constraining it in a way to reproduce the zero anisotropy results for small lengths $L$.

The static potential function we fit in both the anisotropic and transverse directions has the following form
\be\la{vlfitg}
V(L,a,T)=c_1(a/T)+\frac{c_2(a/T)}{L}(1+c_4(a/T) L^2 T^2+ c_3(a/T) L^4 T^4 )~.
\ee
The quantities $L T, ~a/T$ and $a L$ are dimensionless and are the combinations we expect to appear. Moreover due to form of the background metric, we expect the anisotropic parameter to appear in even powers.

Here we solve numerically the potential for different values of low
anisotropy $k_i=a/T$ and find the $c_i$ constants of \eq{vlfitg}.
Then each factor we fit it again to a function of the anisotropic
parameter that we expect from dimensional analysis. There are two
ways to constrain the fitting. One is to fit the formulas directly
to the numerical anisotropic solution and the second one is to use
the isotropic solution for small lengths $L$ and dimensional
analysis to fix the constants that do not depend on the anisotropy.
Both ways as expected give very similar results.

Presenting the results of the second approach we find the following fit to be the best for the anisotropic direction
\bea\nn
2 \pi \alpha' V_{\parallel}(L)&=&18.927 - \frac{1}{L}\prt{1.436+ 20.357 L^4 T^4} + 0.172 \frac{a^2}{T^2} \\
&&-\frac{1}{L}\prt{ 0.0253 \frac{a^2}{T^2} - 67.168 a^2 L^4 T^2 }\la{fitp}
\eea
which means that $c_4(a)$ from \eq{vlfitg} is zero. Similarly the potential to the transverse direction reads
\bea\nn
2 \pi \alpha' V_{\perp}(L)&=&18.927 - \frac{1}{L}\prt{1.436 + 20.3567 L^4 T^4} + 0.124 \frac{a^2}{T^2} \\
&&-\frac{1}{L}\prt{ 0.0246 \frac{a^2}{T^2} - 67.393 a^2 L^4 T^2
}~.\la{fitt} \eea
For the term $a^2/\prt{L T^2}$ we have kept additional decimal indices to the corresponding numerical factor just to show that the numerics indicate different value along the anisotropic and transverse direction.
\begin{figure*}[!ht]
\begin{minipage}[ht]{0.5\textwidth}
\begin{flushleft}
\centerline{\includegraphics[width=70mm]{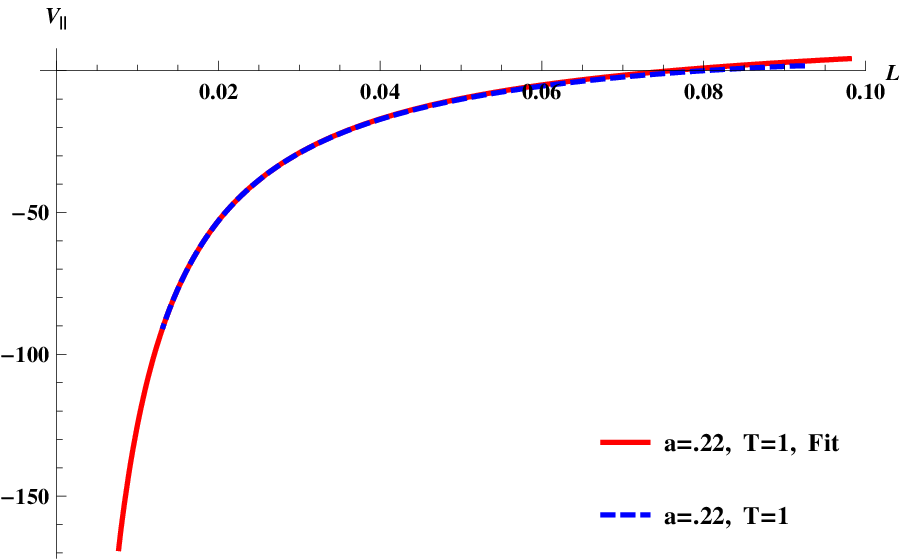}}
\caption{\small{The static potential along the anisotropic direction with the fitting \eq{fitp}.\vspace{.7cm}}}\label{fig:figv1}
\end{flushleft}
\end{minipage}
\hspace{0.3cm}
\begin{minipage}{0.5\textwidth}
\begin{flushleft}
\centerline{\includegraphics[width=70mm]{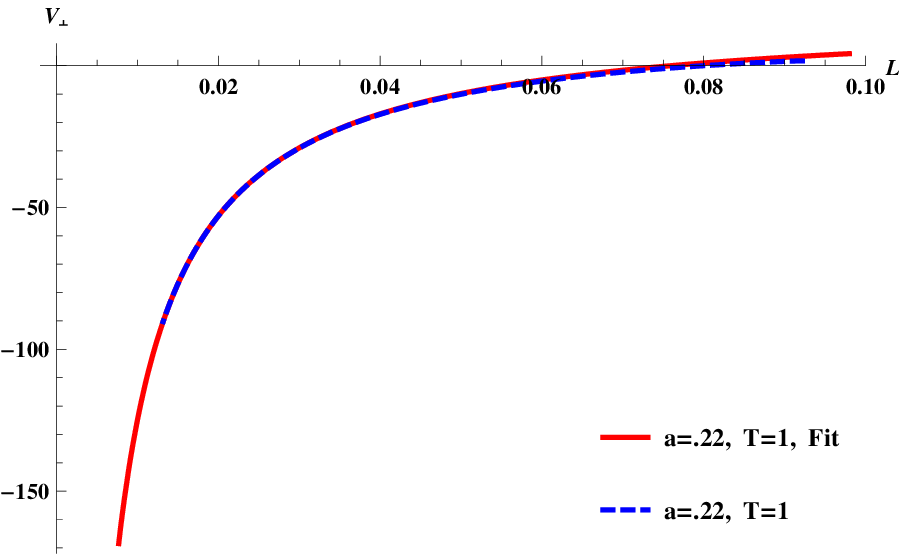}}
\caption{\small{The static potential along the transverse direction with the fitting \eq{fitt}. In both Figures the fitting is very good at least until the critical distance $L_c$.   \vspace{0.cm}}}
\label{fig:figv2}
\end{flushleft}
\end{minipage}
\end{figure*}
In Figures \ref{fig:figv1} and \ref{fig:figv2} we plot the static
potentials along the two directions with the numerical fitting.
Notice that the fitting is very good, this is also indicated by the values of
fitting parameters of the analysis. By giving numerical values for the length, temperature and the anisotropic parameter to equations
\eq{fitp} and \eq{fitt} we find for the absolute values
$V_\parallel<V_\perp<V_{iso}$ for the same length and temperature,
in agreement with the numerical analysis of \cite{Giataganas_aniso} and \cite{Rebhan:2012bw,matqq,Chakraborty:2012dt}.

\subsection{Thermal Width in Anisotropic Theory}

In this section we use the imaginary part of potential in a hot
anisotropic plasma to investigate the thermal width of heavy
quarkonium. Such imaginary contributions to the potential are
related to quarkonium decay processes in the QGP.  It was shown that
the spectral's function peak becomes wider with the temperature and
finally disappears. The width increases with temperature while the
binding energy decreases and exceeding the binding energy already
before the latter disappears and quarkonium melts \cite{Laine:2006ns,Laine:2007gj}.
The effect of the imaginary part of the potential on the thermal widths of the states  anisotropic plasmas has been studied in
\cite{imaginary2,Dumitru:2009ni,Dumitru:2010id}, where it was found that the imaginary part depends on the anisotropy.

We use the anisotropic imaginary potential studied in previous sections to
calculate the thermal width of $\Upsilon$ meson. In section \ref{real-potential} we found semi-analytically the modifications due to anisotropy in the Coulombic part of the real potential which will determine the Coulombic type wavefunction $|\psi>$, needed in equation \eq{gamma}.

There is an ambiguity to calculate the thermal width both in
isotropic and anisotropic background. As we can see in Fig.
\ref{fig:l1} the imaginary potential we found in this approach is
well-defined for a seperation in $(L_0~,~L_m)$. On the other hand
from physical point of view we expect that the imaginary potential
should exist also for larger seperation and this is certainly a weak
point of our current gauge/gravity calculation. However, we can fix the
ambiguity by assuming that the solution we have found in the  $(L_0~,~L_m)$ region extends to larger lengths, and can be found by extrapolating the curve there.

Therefore we do the integration in \eq{gamma}
in $(L_0~,~L_m)$ and then we do the calculation in the region $(L_0~,~\infty)$ by using a reasonable extrapolation for imaginary potential.
For $L<L_0$ the imaginary potential is considered to be zero in both approaches. We extrapolate a straight-line approximation for the imaginary potential as a function of $L$ while the Coulombic wavefunction decreases exponentially with the Bohr radius. 

\begin{figure*}[!ht]
\begin{minipage}[ht]{0.5\textwidth}
\begin{flushleft}
\centerline{\includegraphics[width=3in]{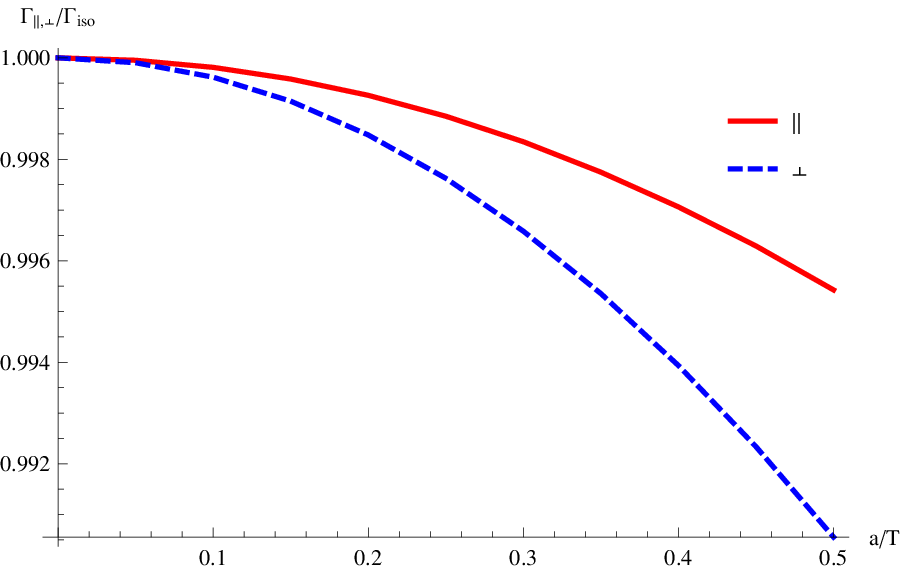}}
\caption{\small{The ratio of thermal width in anisotropic plasma to isotropic depending on the anisotropy for fixed temperature with extrapolation.\vspace{0cm}}}\label{TW1b-plota}
\end{flushleft}
\end{minipage}
\hspace{0.3cm}
\begin{minipage}{0.5\textwidth}
\begin{flushleft}
\centerline{\includegraphics[width=3in]{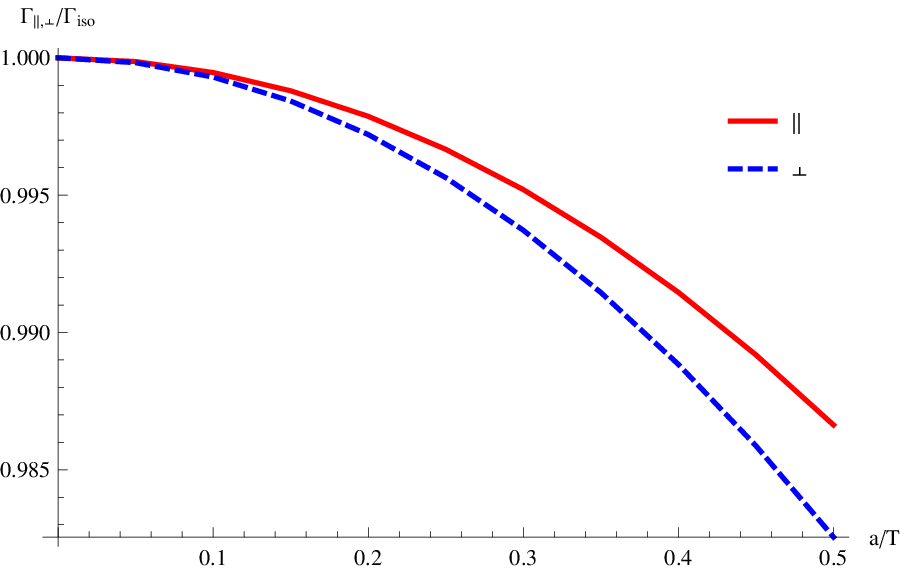}}
\caption{\small{Same quantities as in Figure \ref{TW1b-plota} without extrapolation \vspace{.4cm}.   \vspace{0.2cm}}}
\label{TW1b-plotb}
\end{flushleft}
\end{minipage}
\end{figure*}

\begin{figure*}[!ht]
\begin{minipage}[ht]{0.5\textwidth}
\begin{flushleft}
\centerline{\includegraphics[width=3in]{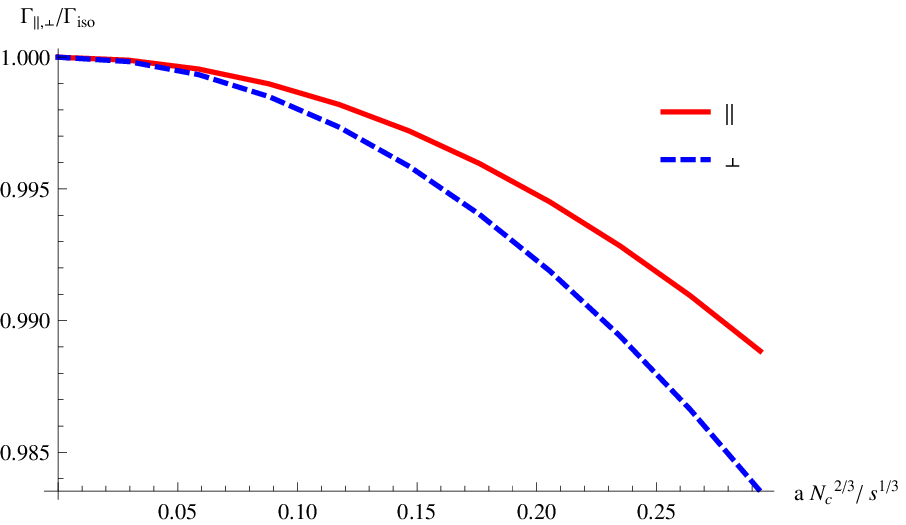}}
\caption{\small{ The ratio of thermal width in anisotropic plasma to isotropic depending on the anisotropy for fixed entropy density with extrapolation.\vspace{0cm}}}\label{TW1b-plotc}
\end{flushleft}
\end{minipage}
\hspace{0.3cm}
\begin{minipage}{0.5\textwidth}
\begin{flushleft}
\centerline{\includegraphics[width=3in]{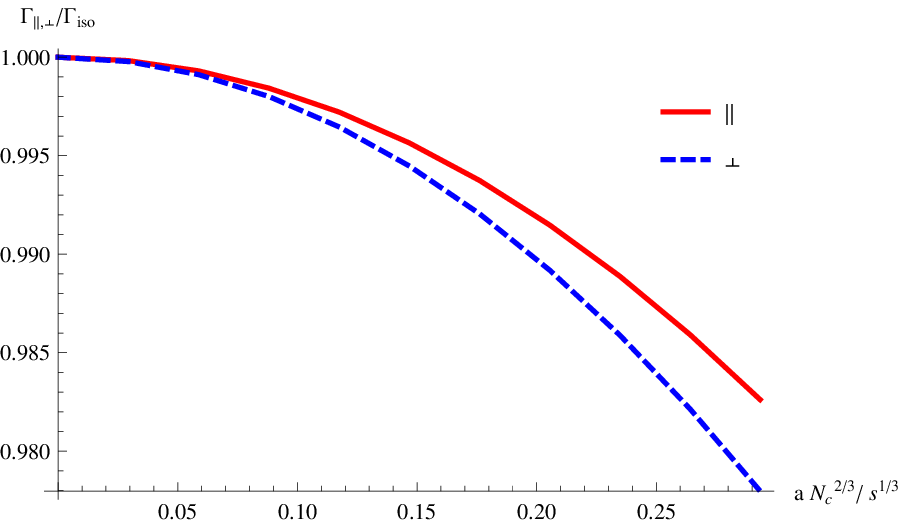}}
\caption{\small{ Same quantities as in Figure \ref{TW1b-plotc} without the extrapolation .\vspace{0.2cm}}}
\label{TW1b-plotd}
\end{flushleft}
\end{minipage}
\end{figure*}

By fixing the energy density we evaluate the thermal widths with and without extrapolation in both comparison schemes and both methods lead to decrease of the thermal width
\be
\Gamma_{\perp}<\Gamma_{\parallel}<\Gamma_{iso}~
\ee
and are presented in Figures \ref{TW1b-plota}, \ref{TW1b-plotb}, \ref{TW1b-plotc} and \ref{TW1b-plotd}.
In order to obtain some naive 'quantitative' results we choose representative values and by fixing the entropy density we get $T_{\text{asYM}}\sim190$MeV. By comparing the renormalized charge of
conformal theories with the lattice data for small separation
lengths of the dipole it has been found that a t'~Hooft coupling of $\l=5.5$
gives a good agreement in the two theories \cite{Gubser:2006qh}. More interesting are the results with the extrapolation were
the values given for the thermal width in the
energy density scheme are around $55$ MeV for $\Upsilon$ in isotropic $\cN=4$ sYM using $m_Q=4.8$ GeV.

\section{Discussion}

In this paper we have extended the imaginary potential in
gauge/gravity dualities introduced in \cite{Noronha:2009da} to the
anisotropic strongly coupled plasma. We considered a long string
with turning point close to the horizon. The fluctuations around the
turning point change the sign of the quantity under the square root of the
Nambu-Goto action and therefore an imaginary part of the action is
generated. We found the imaginary potential for a generic anisotropic
background in terms of the
distance of the dipole. It is interesting that the formula for the
imaginary potential \eq{im-potential2} does not include any
integration although the usual integration in the distance $L$
\eq{staticL1} of the quarks is needed.

We find that in general there is a turning point in the imaginary potential which is similar to the one appears in the real part of the potential in finite temperature. Imposing the boundary and physical conditions on our results, we keep the upper branch of the solution as acceptable from $\prt{L_0,L_m}$ where $L_0$ is the pair distance where the complex potential it is generated and $L_m$ is the turning point. Notice that for larger than $L_m$ distances the real part of the potential suffers from certain instabilities \cite{stability} and becomes problematic. In order to avoid the turning point appearance, there might be a need to modify the current method, for example with a different subtraction scheme of the UV divergences. Notice that our current background satisfies the conditions of \cite{giataganasUV} and therefore even the Legendre subtraction scheme of the infinities can be applied in the anisotropic background successively. So if any modifications are introduced in the cancelation of the UV divergences should be in a consistent way with the Legendre transform. In our case, when we use the potential to find the thermal width we use a simple extrapolation of our current result for lengths larger than $L_m$.

Applying our generic results to the anisotropic theory generated by
a space dependent axion, we find for small and large anisotropies 
that the absolute value of
imaginary part of the potential is increased in presence of
anisotropy opposing the results of the real part of the potential
\cite{Giataganas_aniso} 
for fixed temperature. Stronger enhancement happens along the
anisotropic direction. Further increase of the anisotropy leads to
further enhancement of the imaginary part while its gradient along
the anisotropic direction is much larger than the transverse one.
The fact that the modifications on the observables are stronger
along the anisotropic direction is something that has been observed
before and reflects the stronger dependence of the background
geometry on the anisotropy along the anisotropic direction than to
the transverse one. When the entropy density gets fixed
we observe increase along the anisotropic direction, while along the transverse space we get a decrease of the imaginary potential.

Then we use the imaginary potential to estimate the thermal width in
the anisotropic plasma. We use a numerical fitting to obtain a
function of the real part of the potential using dimensional
analysis. It is used to obtain a Coulombic wavefunction in the
region of small $L$. Then by calculating the
$<\psi|\text{ImV}_{Q\bar{Q}}|\psi>$ we obtain the modifications of
the anisotropy in the thermal width. The physical results come by
extrapolating the imaginary potential function as a continuous
straight line. Then the thermal width is decreased with the
following order $\Gamma_{\perp}<\Gamma_{\parallel}<\Gamma_{iso}$ and
gives values for $\Upsilon$ around $55$ MeV when extrapolation is used and
the entropy density is fixed.\newline

\textbf{Acknowledgments} We would like to thank A.
Dumitru, J. Noronha  and M. Strickland for useful correspondence. The research of D. G. is implemented under the "ARISTEIA" action of the "operational programme education and lifelong learning" and is co-funded by the
European Social Fund (ESF) and National Resources. Part of this work
was also done while D. G. was supported by a Claude Leon postdoctoral
fellowship. H.~S. and K.~B.~F would like to thank the Institute for
Research in Fundamental Sciences (IPM) and the Abdus Salam
International Center for Theoretical Physics (ICTP) for hospitality
during different stages of preparing this project. H. S. is
supported by National Research Foundation (NRF). Any opinion,
findings and conclusions or recommendations expressed in this
material are those of the H. S. and therefore the NRF do not accept
any liability with regard thereto.
\appendix

\section{Q\={Q} Static potential in anisotropic background}\la{app:qq}

In this section we present the calculation of the Wilson loop corresponding to the static potential in an anisotropic
background  \cite{Giataganas_aniso}. We begin with the generic metric of a $d+1$ dimensional background
\be\label{metricqq1}
ds^2=G_{00}d\t^2+G_{ii}dx_i^2+G_{uu}du^2~,\hspace{10mm}i=1,2,\cdots,d-1
\ee
and choose the static gauge for the string which is extended along the radial direction
\be
x_0=\t~,\qquad x_p=\sigma,\qquad\mbox{and}\qquad u=u(\s)~,
\ee
in which $x_p$ could be any spatial dimension $x_i$ and it is the direction along which the pair is aligned.
In the case of anisotropic background we are interested we use $x_{\parallel}$ for the anisotropy direction and $x_{\perp} $ for the transverse directions.
The induced metric for the chosen string configuration reads
\bea
g_{00}=G_{00},\quad\quad g_{11}=G_{pp}+G_{uu}u'^2.
\eea
Working in Lorentzian signature the Nambu-Goto action is
\bea\la{sgener1}
S=\frac{1}{2\pi\a'}\int d\s d\t \sqrt{f+g\,u'^2}=:\frac{1}{2\pi\a'}\int d\s d\t \sqrt{D}~,
\eea
where functions $f$ and $g$ are defined in \eqref{fmg1}
and the corresponding Hamiltonian
\be
H=-\,\frac{ f}{\sqrt{D}}
\ee
 is a constant of motion. Setting it equal to $-c$ we solve for $u'$ to obtain the turning point equation
\be\label{uprime01}
u'=\pm\sqrt{\frac{(f- c^2)\,f}{ c^2\,g}}~,
\ee
which is solved for
\be\label{tpsol1}
G_{uu}^{-1}=0~,\qquad\mbox{or}\qquad G_{pp}=0~,\quad\quad\mbox{or}\qquad f= c^2~.
\ee
Usually the last equation specifies how deep the world-sheet goes into the bulk, ie. the turning point $u_{0}$.

The length of the string in the boundary which corresponds to the distance of the heavy probe quarks is given by
\be\label{staticL1}
L=2\int_{\infty}^{u_{0}}\frac{du}{u'}=2\int_{u_{0}}^{\infty}  du \sqrt{\frac{- G_{uu} c^2}{(G_{00}G_{pp}+ c^2)G_{pp}}}=2\,\int_{u_0}^{\infty}du~\left[\frac{f}{g}\left(\frac{f}{f_0}-1\right)\right]
^{-\frac{1}{2}}~~.
\ee
where the last equation is written using the notation for $f,~g$ introduced in \eq{fmg1} and $f_0=f(u_0)$.
The energy of the string using as renormalization method the mass subtraction of the two free quarks is
\bea\nn
2\pi\a' E&=&2\left(\int_{u_{0}}^{\infty} d\s \cL -\int_{u_{k}}^{\infty} du \sqrt{ g}\right)\nonumber\\\nn
&=& c L+2\left[  \int_{u_{0}}^{\infty} du \sqrt{g}\left(\sqrt{1-\frac{c^2}{f}}-1\right)- \int_{u_{h}}^{u_0} du \sqrt{g}\right]~,\\\label{staticE1}
\eea
where $u_h$ is the horizon of the metric. The factors of two in the right hand side of the above equation comes from the fact that the world-sheet is symmetric with respect to turning point $u_0$.
\section{Imaginary potential in small anisotropy: An analytic approach}
\la{app:imsml}
In this appendix we attempt to find an analytic expression of the form of imaginary potential for low values of $a/T$ in the transverse and parallel directions. Firstly we fit the form of the length $L$ \eq{staticL1} for a small $LT$ to an expected function obtained by dimensional arguments
\bea
L~T=\,\left[b_{1}+b_{2}\,\frac{a^2}{T^2}\right]\,\tilde{u_0}+\mathcal{O}(a^4),
\eea
where for the different directions we have
with
\bea
&& b_{1,\perp}\simeq0.34585,\hspace{10mm}b_{2,\perp}\simeq-0.002684,\\
&&b_{1,\parallel}\simeq0.34585 ,\hspace{10mm}b_{2,\parallel}\simeq-0.001999 ~.\label{aniso-LT}
\eea
Solving the above equation in $u_0$ and substituting to \eq{im-potential2} we obtain a lengthy expression for the static potential. For completeness we write the result


\bea
&&\text{Im}V_{Q\bar{Q}}=\frac{ \sqrt{\lambda } \pi  \left(b_1^4-3 (LT)^4\right)}{24 \sqrt{2} b_1^3 L}+\frac{\sqrt{\lambda} }{3456 \sqrt{2} \pi  b_1^7 \left(b_1^2+(LT)^2\right)^2\,L}\,\frac{a^2}{T^2}\times\nn\\
&&\hspace{10mm}\Bigg[ 6 b_1^{11} \left(b_1+24 \pi^2 b_2\right)-3 b_1^9  \left(b_1-96 \pi^2 b_2\right) (LT)^2+3 b_1^7  \left(480 \pi^2 b_2-7 b_1\right)(LT)^4\nn\\
&&\hspace{10mm}+2 b_1^5 \left(79 b_1+1296 \pi^2 b_2\right) (LT)^6+16 b_1^3 \left(7 b_1+81 \pi^2 b_2\right) (LT)^8-63 b_1^2 (LT)^{10}\nn\\
&&\hspace{10mm}-45 (LT)^{12}-15\log (2) b_1^4 \left(b_1^2+(LT)^2\right)^2  \left(b_1^4+9 (LT)^4\right)\nn\\
&&\hspace{10mm}-3 b_1^4 \left(b_1^2+(LT)^2\right)^2 \left(b_1^4-63 (LT)^4\right) \log \left(\frac{(LT)^2}{b_1^2}+1\right)\nn\\
&&\hspace{10mm}\pm3  \left(b_1^2+(LT)^2\right) \left(5 b_1^8-4 b_1^6 (LT)^2-26 b_1^4 (LT)^4+9 (LT)^8\right)(LT)^2
\Bigg]
\eea
where $+(-)$ sign is for parallel (transverse) direction.


The important information we can obtain from here is that the form of the potential in the region we are looking at is
\be
\text{Im}V_{Q\bar{Q}}=-a_1-a_2 \frac{a^2}{T^2}+\mathcal{O}(a^4)~,
\ee
with $a_1$ and $a_2$ positive numbers and therefore the increase in absolute value is obvious in agreement with the numerical results. The analytical result needs to be taken cautiously since a fitting has been used.

\section{Relation between weakly coupled anisotropic parameters $\xi$ and strongly coupled $a$}\label{we-soa}
In this appendix we derive the qualitative behavior between the anisotropic parameter $\xi$ and the parameter $a$ of our supergravity background thinking along the lines of \cite{Giataganas_aniso,Mateos:2011ix}. By defining $\D$ as
\be
\D:=\frac{P_T}{P_L}-1=\frac{P_{x_1 x_2}}{P_{x_3}}-1~,
\ee
the distribution \eq{faniso} is used to calculate the pressures through the stress energy tensor components. After some algebra the relation to $\xi$ is found to be \cite{0902.3834}
\be\label{dxi1}
\D=\frac{1}{2}(\xi-3)+\xi\left(\left(1+\xi\right)\frac{\arctan\sqrt{\xi}}
{\sqrt{\xi}}-1 \right)^{-1}~.
\ee
In the small $\D$ or $\xi$ limit \eq{dxi1} becomes
\be\la{xil11}
\lim_{\xi\rightarrow 0}\D=\frac{4}{5}\xi +\cO(\xi^2)
\ee
and in the large $\xi$ limit
\be\la{xil22}
\lim_{\xi\rightarrow \infty}\D=\frac{1}{2}\xi +\cO(\sqrt{\xi})~.
\ee
For the anisotropic background in the high temperature limit using the pressures of \eq{ppxx} we get for $\D$
\be\label{da}
\D=\frac{a^2}{2 \pi^2 T^2}~.
\ee
and for $T\gg a\Rightarrow\D\ll 1$  the two anisotropic parameters are related as
\be\label{xiaa1}
\xi\backsimeq \frac{5 a^2}{8 \pi^2 T^2}~.
\ee
It should be mentioned here that the equation \eq{xiaa} only through the pressure anisotropies and the generation of the anisotropy differ, since the anisotropic theory we are using here comes from a position $\theta$ dependent angle. However, it provides a very useful connection between the anisotropic parameters of the different models.


\end{document}